\documentclass[a4paper,11pt]{article}
\usepackage{pos}
\usepackage{slashed}
\usepackage{caption}
\usepackage{subcaption}
\def\ba{\begin{eqnarray}}
\def\ea{\end{eqnarray}}
\def\be{\begin{equation}}
\def\ee{\end{equation}}

\newcommand{\la}{\label}

\def\symff{\text{SYM}_{4,4}}
\def\symod{\text{SYM}_{1,{\cal D}}}
\def\symot{\text{SYM}_{1,10}}
\title{Some recent advances in the understanding of ${\cal N}=4$ supersymmetric Yang-Mills thermodynamics }
\author*[a]{Ubaid Tantary} 
\author[a,b]{Qianqian Du}
\author[a]{Michael Strickland}

\affiliation[a]{Department of Physics, Kent State University,\\
800 E Summit St, Kent, USA}

\affiliation[b]{Institute of Particle Physics and Key Laboratory of Quark and Lepton Physics (MOS), Central China Normal University,\\
430079, Wuhan, China}

\emailAdd{utantary@kent.edu}
\emailAdd{duqianqianstudent@mails.ccnu.edu.cn}
\emailAdd{mstrick6@kent.edu}
\abstract{ The interest in the thermodynamics of supersymmetric Yang-Mills started after Maldacena proposed the duality between string theory on AdS backgrounds and the large-N limit of SYM theories. One of the motivations to study the thermal properties of ${\cal N}=4$ supersymmetric Yang-Mills in four dimensions ($\symff$) is that at high temperatures, the weak-coupling limit of this theory has many similarities with high temperature quantum chromodynamics (QCD). In this proceedings contribution, we review recent calculations of the resummed perturbative free energy of ${\cal N}=4$ supersymmetric Yang-Mills in four spacetime dimensions through second order in the 't Hooft coupling $\lambda$ at finite temperature and zero chemical potential. We compare our final result with prior results obtained in the weak and strong-coupling limits and construct a generalized Pad\'{e} approximant that interpolates between the weak-coupling result and the large-$N_c$ strong-coupling result.  }

\FullConference{%
  * Particles and Nuclei International Conference - PANIC2021 ***\\
  *** 5 - 10 September, 2021 ***\\
  *** Online ***
}

\begin{document}
\maketitle

\section{Introduction}
The perturbative expansion of the free energy of hot non-Abelian gauge theory and in our case $\mathcal{N}=4$ supersymmetric Yang-Mills in four dimensions  with $N_c$ colors and gauge coupling $g$ can be written in the form 

\be \label{highT}
\lim_{\lambda \rightarrow 0} {\cal F} \sim T^4 \big[ a_0 + a_2 \lambda + a_3 \lambda^{3/2} + \big( a_4+ a_4^{\prime} \log  \lambda \big) \lambda^2 + \mathcal{O}(\lambda^{5/2}) \big]  \, ,
\ee
where $\lambda=g^2 N_c$ is the 't Hooft coupling. The leading term in this expression is the free energy of an ideal plasma and the ${\cal O}(\lambda)$ correction can be obtained by computing two-loop Feynman diagrams. The next contribution is ${\cal O}(\lambda^2)$ and comes from three-loop contributions. However, a problem emerges because one finds uncanceled infrared divergences at the three-loop level if one uses  bare propagators. This happens in QCD as well and there  the infrared divergences can be eliminated by summing over the so-called ring diagrams~\cite{1}. The solution is similar in $\symff$ and the only difference with QCD is the number and types of degrees of freedom. In the weak-coupling limit, the free energy of $\symff$ has been calculated through order $\lambda^{3/2}$ in ~\cite{2} and in the opposite limit of strong coupling, the behavior of the $\symff$ free energy was studied using the AdS/CFT correspondence in ~\cite{3}.
\section{$\mathcal{N}=4$ supersymmetric Yang-Mills theory in $4$-dimensions ($\symff$)}
The $\symff$ theory can be obtained by dimensional reduction of $\symod$ in \mbox{$\mathcal{D}=\mathcal{D}_{\textrm{max}}=10$} with all fields being in the adjoint representation of $SU(N_c)$. The Lagrangian that generates the perturbative expansion for $\symff$ in Minkowski-space can be expressed as
\ba
&& \mathcal{L}_{\symff} = \textrm{Tr}\bigg[{-}\frac{1}{2}G_{\mu\nu}^2+(D_\mu\Phi_A)^2+i\bar{\psi}_i { D}\psi_i-\frac{1}{2}g^2(i[\Phi_A,\Phi_B])^2 \nonumber \\
&& \hspace{2.5cm} - i g \bar{\psi}_i\big[\alpha_{ij}^{\texttt{p}} X_{\texttt{p}}+i \beta_{ij}^{\texttt{q}}\gamma_5Y_{\texttt{q}},\psi_j\big] \bigg] +\mathcal{L}_{\textrm{gf}}+\mathcal{L}_{\textrm{gh}}+\Delta\mathcal{L}_{\textrm{SYM}} \, ,
\ea
  with $\Phi_A \in (X_1,Y_1,X_2,Y_2,X_3,Y_3)$  and $X_{\texttt{p}}$ and $Y_{\texttt{q}}$ denote scalars and pseudoscalar fields, respectively.\\

We are in general interested in supersymmetric field theories with supercharges in dimensions $D\leq \mathcal{D}_{\textrm{max}}$, with $D$ being an integer. The evaluation of Feynman diagrams for theories that are obtained by dimensional reduction of $\symod$ can be carried out in a simple way that preserves the supersymmetry by taking all fields  to be $\mathcal{D}$-dimensional tensors or spinors and all momentum to be $d=D-2\epsilon$ vectors. This scheme was introduced by W. Siegel and is called  regularization by dimension reduction (RDR)~\cite{4} .\\
    \section{Resummation in $\symff$}   
     Since we want to obtain the thermodynamic functions up to $\mathcal{O}(\lambda^{2})$, we need to calculate Feynman diagrams through three loop order. However, at three loop level in QCD ~\cite{1}, infrared divergences appear that need to be canceled by summing over the ring diagrams appearing in the thermal mass counterterm.  As detailed in ref.~\cite{1}, in order to systematically resum the necessary diagrams, we need to modify the static bosonic propagators by incorporating gluon and scalar thermal masses, $m_D$ and $M$, respectively.  

Following Arnold and Zhai, we introduce thermal masses, $m_D$ and $M$, only for the zero Matsubara modes of the gluon and scalar fields.  The resulting reorganized Lagrangian density in frequency space can be rewritten as
\ba\la{lagresum}
\mathcal{L}_{\symff}^{\textrm{resum}}&=& \{\mathcal{L}_{\symff}+ \textrm{Tr}[m_D^2 A_0^2\delta_{p_0} -M^2\Phi_A^2\delta_{p_0}]\} 
-  \textrm{Tr}[m_D^2 A_0^2\delta_{p_0} -M^2\Phi_A^2\delta_{p_0}]\, ,
\ea
Then we absorb the two $A_0^2$ and $\Phi^2 $ terms in the curly brackets into our unperturbed Lagrangian $\mathcal{L}_0$, and treat the two terms outside the curly brackets as a perturbation.\\
\begin{figure}[h!]
     \centering
     \begin{subfigure}[b]{0.4\textwidth}
         \centering
         \includegraphics[width=\textwidth]{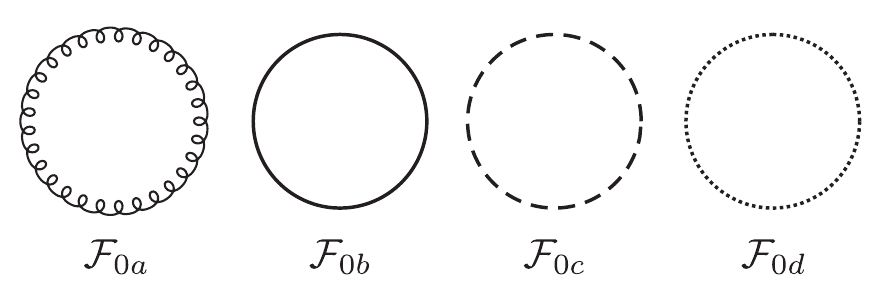}
         \caption{One-loop diagrams contributing to the $\symff$ free energy}
         \label{fig:oneloop}
     \end{subfigure}
     $\;\;\;\;$
     \begin{subfigure}[b]{0.5\textwidth}
         \centering
         \includegraphics[width=\textwidth]{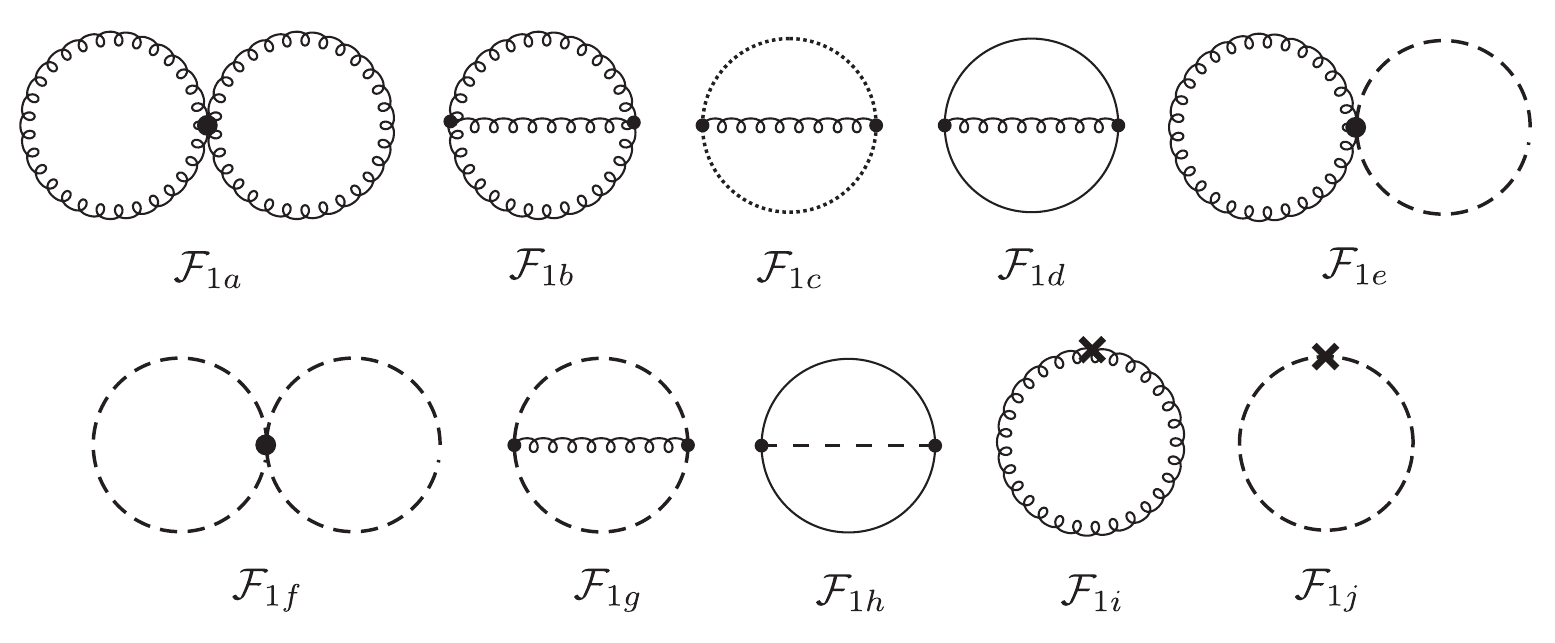}
         \caption{Two-loop contributions to the  $\symff$ free energy. The crosses are the thermal counterterms produced by the last two terms of (\ref{lagresum}).}
         \label{fig:three sin x}
     \end{subfigure}
     \hfill
     \begin{subfigure}[b]{0.4\textwidth}
         \centering
         \includegraphics[width=\textwidth]{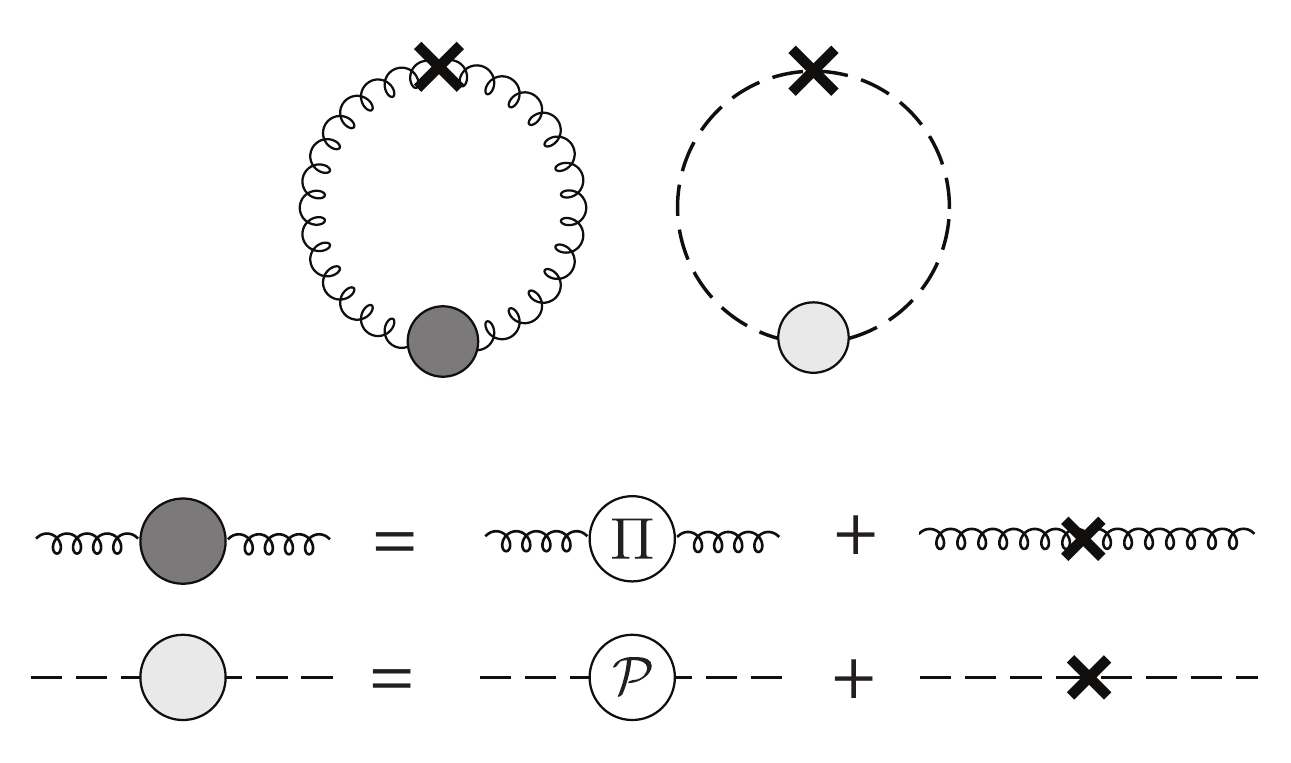}
         \caption{Three-loop gluon and scalar counterterm diagrams in $\symff$}
         \label{fig:y equals x}
     \end{subfigure}
     $\;\;\;\;$
     \begin{subfigure}[b]{0.5\textwidth}
         \centering
         \includegraphics[width=\textwidth]{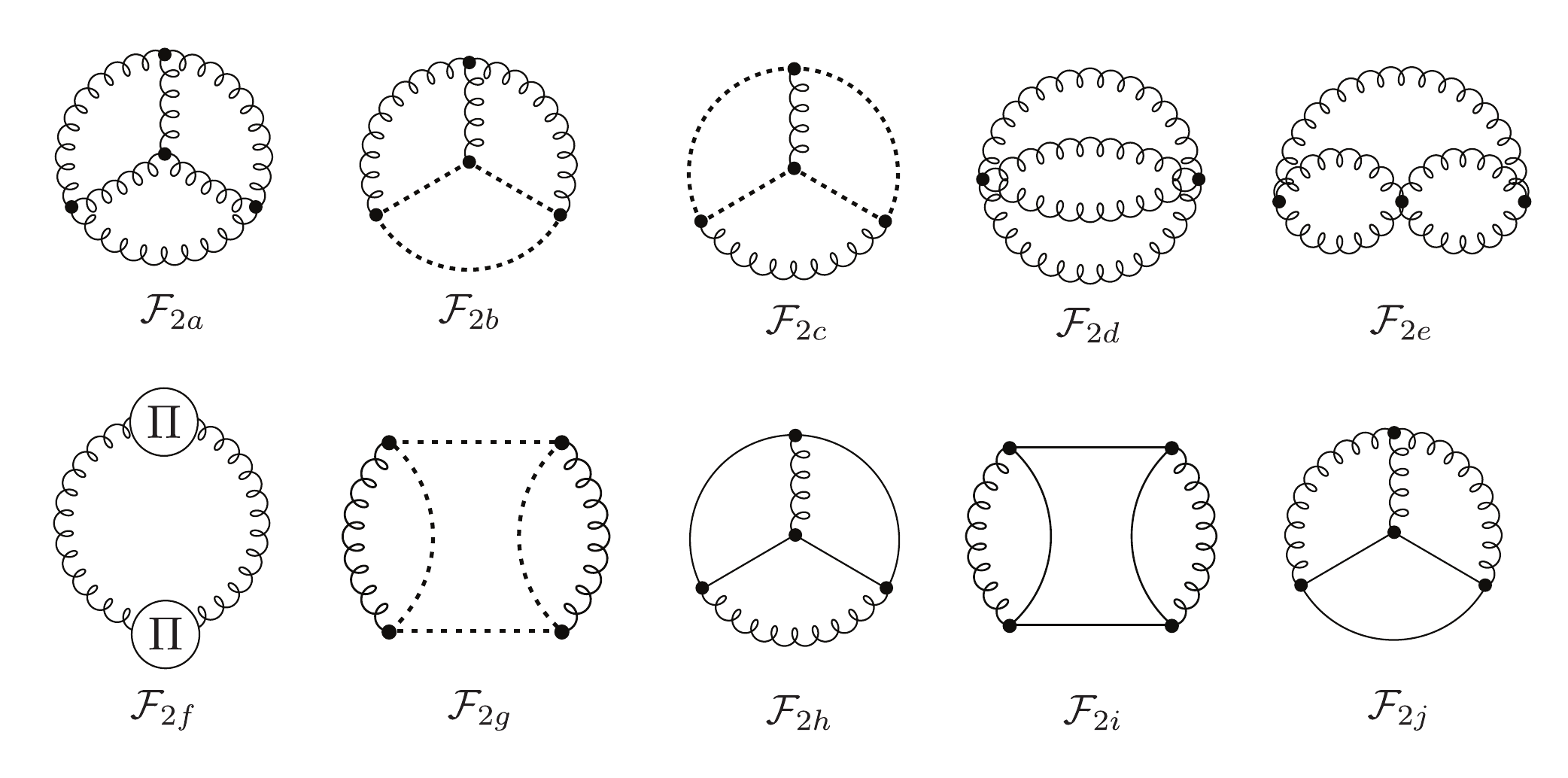}
         \caption{Three-loop vacuum diagrams contributing to the $\symot$ free energy}
         \label{fig:five over x}
     \end{subfigure}
     \hfill        
     \caption{Feynman diagrams up to 3-loop order. The dashed lines indicate a scalar field and dotted lines indicate a ghost field. The crosses are the thermal counter-terms.
}
        \label{fig:three graphs}
\end{figure}
\subsection{The resummed one-loop free energy}
The resummed one-loop free energy can be written as 
\be
F_\text{1-loop}^{\textrm{resum}} = d_A  \mathcal{F}_{0a} +d_F\mathcal{F}_{0b}+d_S\mathcal{F}_{0c}+d_A \mathcal{F}_{0d} \, ,
\ee
with $d_F=4d_A$ and $d_S=6d_A$. By using resummed gluonic and scalar propagators, imposing $D=4$, $m_D^2=2\lambda T^2$, $M^2=\lambda T^2$, and truncating at $\mathcal{O}(\epsilon^0)$ one obtains\be\la{1loopresum}
F_\text{1-loop}^{\textrm{resum}} =-d_A \bigg(\frac{\pi^2 T^4 }{6}\bigg) \bigg[ 1 +\frac{3+\sqrt{2}}{\pi^3} \lambda^{3/2} \bigg]  \, .
\ee
\subsection{The resummed two-loop free energy}
The $\symff$ two-loop free energy can be written as
\be
F_\text{2-loop}^{\textrm{resum}} = d_A \bigg\{\lambda [{\cal F}_{1a}+{\cal F}_{1b}+{\cal F}_{1c}+{\cal F}_{1d}+{\cal F}_{1e}+{\cal F}_{1f}+{\cal F}_{1g}+{\cal F}_{1h}]+{\cal F}_{1i}+{\cal F}_{1j}  \bigg\} \, .
\ee
By using resummed gluonic and scalar propagators one obtains 
\ba\la{2loopresum}
F_\text{2-loop}^{\textrm{resum}} & = & -d_A \bigg(\frac{\pi^2 T^4}{6}\bigg) \bigg[ -\frac{3}{2\pi^2} \lambda-\frac{3}{2\pi^4} \bigg(  \frac{23}{8}+\frac{3\sqrt{2}}{4} +\frac{15 \log 2}{4} -  \log \lambda\bigg)\lambda^2 \bigg]\,.
\ea
\subsection{The resummed three-loop free energy}
The calculation of the massless three-loop vacuum Feynman diagrams in $\symff$ can be accomplished more simply in the corresponding $\symot$ theory.  As a result of this equivalence, one can consider the much smaller set of $\symot$ graphs presented in fig.~1(d), which are topologically equivalent to three-loop QCD vacuum graphs.  The three-loop results in $\symff$  can be obtained by imposing $\mathcal{D} = \mathcal{D}_{\textrm{max}}=10$, $d=4-2\epsilon$ in the $\symod$ theory.
\be
F_\text{3-loop}^{\textrm{vacuum}} = d_A \lambda^2 [{\cal F}_{2a}+ {\cal F}_{2b}+{\cal F}_{2c}+{\cal F}_{2d}+{\cal F}_{2e}+{\cal F}_{2f}+{\cal F}_{2g} +{\cal F}_{2h}+{\cal F}_{2i} +{\cal F}_{2j}]|_{d=4-2\epsilon}^{{\cal D}=10}\,.
\ee
Infrared divergences  are generated in eq.~(8) due to 3-momentum integrations. These divergences are canceled by thermal mass counterterm diagrams in fig.~1(c).\ba\la{3loopresum}
F_\text{3-loop}^{\textrm{resum}} &=& {\cal F}_\text{3-loop}^{\textrm{vacuum}}+\mathcal{F}_{\textrm{3-loop}}^{\textrm{sct}}+\mathcal{F}_{\textrm{3-loop}}^{\textrm{bct}} \nonumber\\& =& -d_A \bigg(\frac{\pi^2 T^4}{6} \bigg) \frac{\lambda^2}{2\pi^4} \bigg[  \frac{27}{8} +3\gamma+3 \frac{\zeta'(-1)}{\zeta(-1)} +5\log 2-6\log\pi \bigg]\,.
\ea
\section{$\symff$ thermodynamic functions to ${\cal O}(\lambda^2)$}\la{thermodynamics}
Combining eqs.~(\ref{1loopresum}), (\ref{2loopresum}), and (\ref{3loopresum}), we obtain our final 
 result for the resummed free energy in the RDR scheme through ${\cal O}(\lambda^2)$.\footnote{We have noticed a small error in Ref.~\cite{5}, recently. The first term on the second line of (10) should be $-{21\over8}$ instead of $-{45\over16}$ as obtained in Ref.~\cite{5}.   Equation \eqref{eq:finalF} already includes this correction.}
\ba
{\cal F} &=& -d_A \bigg(\frac{\pi^2 T^4}{6}\bigg) \bigg\{ 1-\frac{3}{2} \frac{\lambda}{\pi^2} +  \left( 3+\sqrt{2} \right) \left(\frac{\lambda}{\pi^2}\right)^{3/2}  \nonumber \\ && \hspace{1cm} + \bigg[ -\frac{21}{8} -\frac{9\sqrt{2}}{8} + \frac{3}{2} \gamma_E + \frac{3}{2}\frac{\zeta'(-1)}{\zeta(-1)}-\frac{25}{8} \log 2 + \frac{3}{2}\log \frac{\lambda}{\pi^2} \bigg] \left(\frac{\lambda}{\pi^2}\right)^2  \bigg \}. \hspace{8mm} 
\label{eq:finalF}
\ea
\section{Conclusions and Outlook}
In this work, we reviewed the computation of the  thermodynamic function of $\symff$ to ${\cal O}(\lambda^2)$.  The final result, presented in eq.~\eqref{eq:finalF}, extends our knowledge of weak-coupling $\symff$ thermodynamics to include terms at ${\cal O}(\lambda^2)$  and ${\cal O}(\lambda^2 \log\lambda)$.  With the ${\cal O}(\lambda^2)$  and ${\cal O}(\lambda^2 \log\lambda)$ coefficients in the $\symff$ free energy, we then constructed a large-$N_c$ Pad\'{e} approximant that interpolates between the weak- and strong-coupling limits. Figure 2 summarizes our findings.
\begin{figure}[h!]
  \centering
  \includegraphics[width=0.5\textwidth]{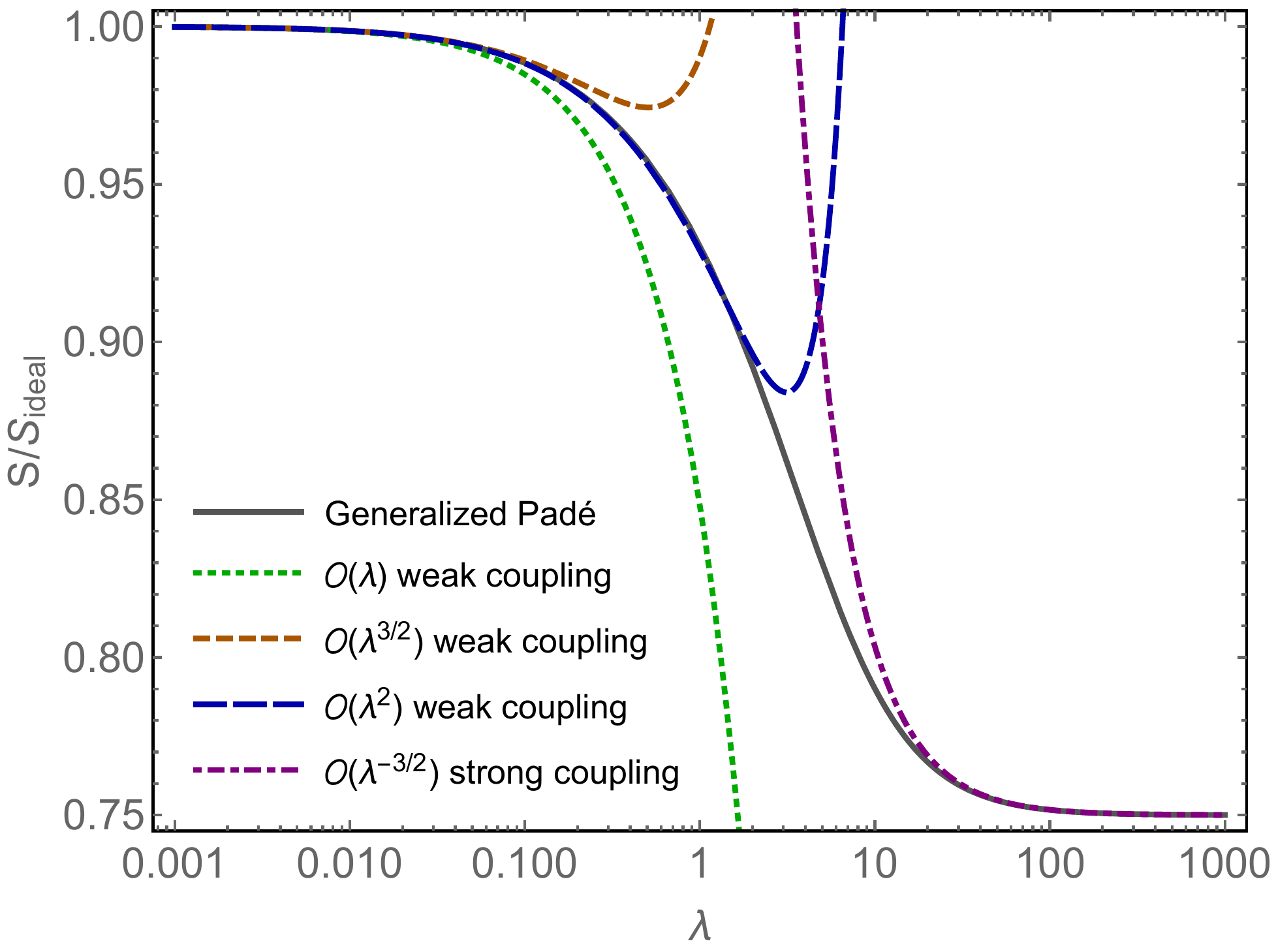}
  \caption{The entropy density ${\cal S}$ normalized by the ${\cal S}_{\rm ideal}$
in ${\rm \symff}$ as a function of the t 'Hooft coupling $\lambda$. }
\end{figure}

We have recently rederived the final result \eqref{eq:finalF} using effective field theory techniques~\cite{6}. We are also working on computing  the coefficient of $\lambda^{5/2}$ in the $\symff$ free energy using effective field theory methods. Finally, we also plan to pursue a three-loop HTLpt calculation of $\symff$ thermodynamics, extending our prior two-loop HTLpt results~\cite{7,8}.
\section*{Acknowledgements}
 Q.D., M.S., and U.T. were supported by the U.S. Department of Energy, Office of Science, Office of Nuclear Physics under Award No.~DE-SC0013470. In addition, Q.D. was supported by the China Scholarship Council under Project No.~201906770021, the National Natural Science Foundation of China Project No.~11935007, and the Guangdong Major Project of Basic and Applied Basic Research No. 2020B0301030008.

\end{document}